\documentclass[aps,prm,twocolumn,showpacs,byrevtex]{revtex4-2}
\let\oldhat\hat
\renewcommand{\hat}[1]{\oldhat{\mathbf{#1}}}

\usepackage{times,mathptm}
\usepackage[dvips]{graphicx,color}

\begin{document}
\title{Formation mechanism of chemically precompressed hydrogen clathrates in metal superhydrides}
\author{Shichang Yao, Chongze Wang, Shuyuan Liu, Hyunsoo Jeon, and Jun-Hyung Cho$^*$}
\affiliation{Department of Physics, Research Institute for Natural Science, and Institute for High Pressure at Hanyang University, Hanyang University,  222 Wangsimni-ro, Seongdong-Ku, Seoul 04763, Republic of Korea}
\date{\today}

\begin{abstract}
Recently, the experimental discovery of high-$T_c$ superconductivity in compressed hydrides H$_3$S and LaH$_{10}$ at megabar pressures has triggered searches for various superconducting superhydrides. It was experimentally observed that thorium hydrides, ThH$_{10}$ and ThH$_9$, are stabilized at much lower pressures compared to LaH$_{10}$. Based on first-principles density-functional theory calculations, we reveal that the isolated Th frameworks of ThH$_{10}$ and ThH$_9$ have relatively more excess electrons in interstitial regions than the La framework of LaH$_{10}$. Such interstitial excess electrons easily participate in the formation of anionic H cage surrounding metal atom. The resulting Coulomb attraction between cationic Th atoms and anionic H cages is estimated to be stronger than the corresponding one of LaH$_{10}$, thereby giving rise to larger chemical precompressions in ThH$_{10}$ and ThH$_9$. Such a formation mechanism of H clathrates can also be applied to another experimentally synthesized superhydride CeH$_9$, confirming the experimental evidence that the chemical precompression in CeH$_9$ is larger than that in LaH$_{10}$. Our findings demonstrate that interstitial excess electrons in the isolated metal frameworks of high-pressure superhydrides play an important role in generating the chemical precompression of H clathrates.
\end{abstract}
\maketitle

\section{INTRODUCTION}

In recent years, hydrides have attracted much attention theoretically and experimentally because of their promising possibility for the realization of room-temperature superconductivity (SC)~\cite{review-Zurek,review-Eremets}. Motivated by the pioneering idea of Neil Ashcroft on high-temperature SC in metallic hydrogen~\cite{MetalicH_Ashc} and the incessant theoretical predictions of high superconducting transition temperature $T_c$ in a number of hydrides~\cite{LiHx-PANS2009,LiHx-acta cryst.2014,KHx-JPCC2012,CaH6-PANS2012,H3S-Theory,H3S-Theory2,MgH6-RSC-Adv.2015,rare-earth-hydride-PRL2017,rare-earth-hydride-PANS2017,Li2MgH16-PRL2019,HfH10-PRL2020, LaH10-PRB2019-Liangliang,LaH10-PRB2020-Chongze,LaH10-PRB2020-isotope,YH10-Boeri,Compressed-hydrides-Pickett,LaH10-Papa}, experiments have confirmed that sulfur hydride H$_3$S and lanthanum hydride LaH$_{10}$ exhibit $T_{\rm c}$ around 203 K at ${\sim}$155 GPa~\cite{ExpH3S-Nature2015} and 250$-$260 K at ${\sim}$170 GPa~\cite{ExpLaH10-PRL2019, ExpLaH10-Nature2019}, respectively. Subsequently, such a conventional Bardeen-Cooper-Schrieffer type SC has also been experimentally observed in various compressed hydrides at high pressures~\cite{ExpCeH9-Nat.Commun2019T.Cui,ExpCeH9-Nat.Commun2019-J.F.Lin,ExpCeH9-arXiv2021,ExpThH10-Materials Today2020-Oganov,ExpYH6-arXiv2019,Exp-CSH-Nature2020,PrH9}. For examples, ThH$_{10}$ (ThH$_9$) exhibits $T_c$ = 159$-$161 (146) K between 170 and 175 GPa~\cite{ExpThH10-Materials Today2020-Oganov}, while CeH$_9$ exhibits a $T_c$ of ${\sim}$100 K at 130 GPa~\cite{ExpCeH9-arXiv2021}. More recently, carbonaceous sulfur hydride was observed to exhibit a room-temperature SC with a $T_{\rm c}$ of 288 K at ${\sim}$267 GPa~\cite{Exp-CSH-Nature2020}. Therefore, the experimental observations of high-temperature SC in either surfur-containing hydrides~\cite{Exp-CSH-Nature2020,ExpH3S-Nature2015} or superhydrides containing an abnormally large amount of hydrogen~\cite{ExpLaH10-PRL2019,ExpLaH10-Nature2019,ExpCeH9-Nat.Commun2019T.Cui,ExpCeH9-Nat.Commun2019-J.F.Lin,ExpCeH9-arXiv2021,ExpYH6-arXiv2019,ExpThH10-Materials Today2020-Oganov} has launched a new era of high-$T_{\rm c}$ superconductors.

Compared to the existence of metallic hydrogen at high pressures over ${\sim}$400 GPa~\cite{MetalicH1,MetalicH2}, the syntheses of superhydrides with H-rich clathrate structures have been achieved at relatively much lower pressures, because H atoms can be “chemically precompressed” by chemical forces between metal atoms and H cages~\cite{precompress-Ashcroft}. Using density-functional theory (DFT) calculations, the high-$T_c$ superconducting phases of various superhydrides have been predicted to be metastable at higher pressures than a critical pressure $P_c$~\cite{LaH10-Angew2018-Hemley,LaH10-PRB2018-Hemley,LaH10-PRB2019-Chongze,LaH10-PRB2020-Shipley}. It is noticeable that the magnitude of $P_c$ reflects the strength of chemical precompression in superhydrides. Experimentally, the $P_c$ value of ThH$_{10}$ having a fcc-Th framework [see Fig. 1(a)] was measured to be ${\sim}$85 GPa~\cite{ExpThH10-Materials Today2020-Oganov}, much lower than $P_c$ ${\approx}$ 170 of an isostructural superhydride LaH$_{10}$~\cite{ExpLaH10-PRL2019, ExpLaH10-Nature2019}. Furthermore, ThH$_9$ (CeH$_9$) having a hcp metal framework [see Fig. 1(b)] was observed to exhibit a $P_c$ of ${\sim}$86 (80) GPa~\cite{ExpThH10-Materials Today2020-Oganov,ExpCeH9-Nat.Commun2019T.Cui,ExpCeH9-Nat.Commun2019-J.F.Lin,ExpCeH9-arXiv2021}. Based on these existing experimental data~\cite{ExpLaH10-PRL2019, ExpLaH10-Nature2019,ExpThH10-Materials Today2020-Oganov}, it is most likely that $P_c$ changes with respect to metal species: i.e., Group-4 metal hydrides ThH$_{10}$, ThH$_9$, and CeH$_9$ with occupied $f$-subshell electrons have lower $P_c$ values or larger chemical precompressions compared to a Group-3 metal hydride LaH$_{10}$. Our recent DFT calculations for LaH$_{10}$~\cite{seho} revealed that the isolated La framework without H atoms behaves as an electride at high pressures, where some electrons detached from La atoms are well localized in interstitial regions. These interstitial excess electrons are easily captured to H atoms, forming a H clathrate structure in LaH$_{10}$. In the present study, such an electride feature in the La framework of LaH$_{10}$ is compared with other metal frameworks of the above-mentioned superhydrides ThH$_{10}$, ThH$_9$, and CeH$_9$. By the estimation of Coulomb attractions between metal atoms and H cages, we provide an explanation for the different chemical precompressions observed in such superhydrides~\cite{ExpLaH10-PRL2019, ExpLaH10-Nature2019,ExpThH10-Materials Today2020-Oganov,ExpCeH9-Nat.Commun2019T.Cui,ExpCeH9-Nat.Commun2019-J.F.Lin,ExpCeH9-arXiv2021}, as will be discussed below.

In this paper, using first-principles DFT calculations, we perform a comparative study of chemical precompressions in ThH$_{10}$, ThH$_9$, CeH$_9$, and LaH$_{10}$. We find that the isolated metal frameworks of ThH$_{10}$, ThH$_9$, and CeH$_9$ possess more interstitial excess electrons than that of LaH$_{10}$ at an equal pressure of 300 GPa. Such loosely bound electrons can be easily captured to form H clathrate structures with attractive Coulomb interactions between cationic metal atoms and anionic H cages. Using the calculated Bader charges~\cite{Bader} and positions of metal and H atoms in each superhydride, we estimate a chemical pressure acting on H cage around a metal atom. As a result, ThH$_{10}$, ThH$_9$, and CeH$_9$ are found to have larger chemical precompressions than LaH$_{10}$, consistent with the experimentally observed $P_c$ values in these superhydrides~\cite{ExpLaH10-PRL2019, ExpLaH10-Nature2019,ExpThH10-Materials Today2020-Oganov,ExpCeH9-Nat.Commun2019T.Cui,ExpCeH9-Nat.Commun2019-J.F.Lin,ExpCeH9-arXiv2021}. It is thus demonstrated that Group-4 metal hydrides occupying $f$ electrons can be more chemically precompressed compared to Group-3 metal hydride, thereby contributing to lower $P_c$. The present findings illuminate that interstitial excess electrons in the metal frameworks of superhydrides are of importance to generate the chemical precompression of H cages around metal atoms.

\begin{figure}[h!t]
\includegraphics[width=8cm]{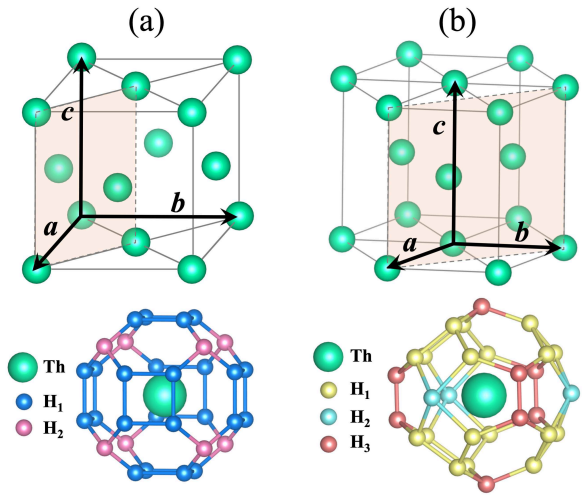}
\caption{Optimized structures of (a) ThH$_{10}$ and (b) ThH$_{9}$ at 300 GPa. ThH$_{10}$ (ThH$_{9}$) has the fcc (hcp) Th framework with the H$_{32}$ (H$_{29}$) cage surrounding each Th atom. There are two (three) different types of H atoms: i.e., H$_{1}$ and H$_{2}$ for ThH$_{10}$ (H$_{1}$, H$_{2}$, and H$_{3}$ for ThH$_{9}$). The (110) planes are drawn in the fcc and hcp lattices.}
\label{figure:1}
\end{figure}

\section{Calculational Methods}

Our first-principles DFT calculations were performed using the Vienna {\it ab initio} simulation package with the projector-augmented wave method~\cite{vasp1,vasp2,paw}. Here, we treated Th 6$s^2$6$p^6$5$f^1$6$d^1$7$s^2$, Ce 5$s^2$5$p^6$4$f^1$5$d^1$6$s^2$, La 5$s^2$5$p^6$5$d^1$6$s^2$ and H 1$s^1$ as valence electrons, including 6$s^2$6$p^6$, 5$s^2$5$p^6$, and 5$s^2$5$p^6$ semicore electrons for Th, Ce, and La, respectively. For the exchange-correlation energy, we employed the generalized-gradient approximation functional of Perdew-Burke-Ernzerhof~\cite{pbe,Th-PRB}. A plane-wave basis was used with a kinetic energy cutoff of 500 eV for ThH$_{10}$ and ThH$_{9}$. The ${\bf k}$-space integration was done with 24${\times}$24${\times}$24 and 18${\times}$18${\times}$12 $k$ points for ThH$_{10}$ and ThH$_{9}$, respectively. All atoms were allowed to relax along the calculated forces until all the residual force components were less than 0.005 eV/{\AA}. We calculated phonon frequencies with the 6${\times}$6${\times}$6 (5${\times}$5${\times}$3) $q$ points for ThH$_{10}$ (ThH$_{9}$) using the QUANTUM ESPRESSO package~\cite{QE}. For CeH$_9$ and LaH$_{10}$, we chose the calculation parameters used in our previous works~\cite{seho,hyunsoo}.

\section{Results}

We first optimize the structures of experimentally synthesized superhydrides ThH$_{10}$, ThH$_9$, CeH$_9$, and LaH$_{10}$ as a function of pressure using DFT calculations. These superhydrides have hydrogen sodalitelike clathrate structures with high-symmetry space groups: i.e., $Fm$$\overline{3}m$ (No. 225) for ThH$_{10}$ and LaH$_{10}$, while P6$_3$/$mmc$ (No. 194) for ThH$_9$ and CeH$_9$. As shown in Fig. 1(a), ThH$_{10}$ (LaH$_{10}$) is constituted by the fcc metal framework, where each Th (La) atom is surrounded by the H$_{32}$ cage consisting of 32 H atoms. Meanwhile, ThH$_9$ (CeH$_9$) have the hcp metal framework with the H$_{29}$ cage surrounding a Th (Ce) atom [see Fig. 1(b)]. Note that there are two (three) species of H atoms composing the H$_{32}$ (H$_{29}$) cages in ThH$_{10}$ and LaH$_{10}$ (ThH$_9$ and CeH$_9$). The optimized structures of these superhydrides show that the lattice constants decrease monotonously with increasing pressure (see Fig. S1 in the Supplemental Material). Accordingly, the averaged bond lengths $d_{\rm M-H}$ between metal and H atoms decrease with increasing pressure [see Fig. 2(a)]. We find that $d_{\rm M-H}$ for ThH$_{10}$ having the H$_{32}$ cage is longer than ThH$_9$ having the H$_{29}$ cage at a given pressure. Meanwhile, $d_{\rm M-H}$ for CeH$_9$ is shorter than that for ThH$_9$, possibly because of the smaller size of Ce atom with the atomic number of $Z$ = 58 compared to Th atom with $Z$ = 90. Interestingly, despite the larger atomic number of Th than La ($Z$ = 57), the $d_{\rm M-H}$ values for ThH$_{10}$ and ThH$_9$ are close to that for LaH$_{10}$ at a given pressure, implying that the former superhydrides have larger chemical precompressions than the latter one. We note that the charges of cationic metal and anionic H atoms are also essential ingredients for determining chemical precompression, as discussed below. Figure 2(b) displays the averaged H$-$H bond lengths $d_{\rm H-H}$ for ThH$_{10}$, ThH$_9$, CeH$_9$, and LaH$_{10}$, which also decrease monotonously with increasing pressure. It is seen that the $d_{\rm H-H}$ values for ThH$_{10}$ and LaH$_{10}$ are shorter than those for ThH$_9$ and CeH$_9$, indicating that the H$_{32}$ cages composed of larger number of H atoms give rise to shorter $d_{\rm H-H}$ compared to the H$_{29}$ cages.

\begin{figure}[h!t]
\includegraphics[width=8.5cm]{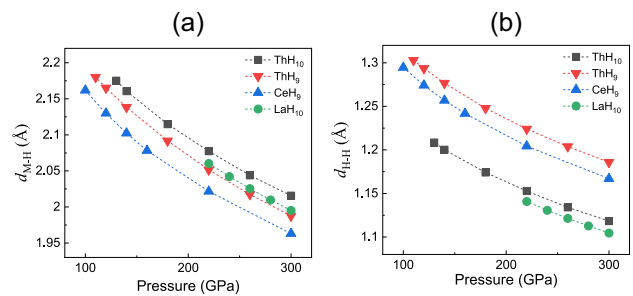}
\caption{Calculated averaged bond lengths (a) $d_{\rm M-H}$ between metal and H atoms and (b) $d_{\rm H-H}$ between H atoms for ThH$_{10}$, ThH$_{9}$, CeH$_{9}$, and LaH$_{10}$ as a function of pressure.}
\label{figure:2}
\end{figure}

In order to examine the dynamical stability of ThH$_{10}$, ThH$_9$, CeH$_9$, and LaH$_{10}$, we calculate their phonon spectra as a function of pressure. As shown in Fig. 3(a), the phonon spectrum of ThH$_{10}$, calculated at 130 GPa, exhibits the softening of low-energy phonon modes (marked by arrows) along the $\Gamma-L$ and $\Gamma-K$ lines. Such H-derived phonon modes finally have imaginary frequencies at 120 GPa [see Fig. 3(b)]. This indicates that the fcc-ThH$_{10}$ phase becomes unstable as pressure decreases. Therefore, the phonon spectra as a function of pressure allow us to predict the $P_c$ values of about 130, 110, 100, and 220 GPa for ThH$_{10}$, ThH$_9$, CeH$_9$~\cite{hyunsoo}, and LaH$_{10}$~\cite{LaH10-PRB2019-Chongze}, respectively (see Fig. S2 in the Supplemental Material). We find that ThH$_{10}$ and ThH$_9$ have much lower $P_c$ than LaH$_{10}$, while their $P_c$ values are close to that of CeH$_9$. The overall trend of these predicted $P_c$ values in four superhydrides are well consistent with the experimentally measured ones of about 85, 86, 80, and 170 GPa for ThH$_{10}$~\cite{ExpThH10-Materials Today2020-Oganov}, ThH$_9$~\cite{ExpThH10-Materials Today2020-Oganov}, CeH$_9$~\cite{ExpCeH9-Nat.Commun2019T.Cui,ExpCeH9-Nat.Commun2019-J.F.Lin,ExpCeH9-arXiv2021}, and LaH$_{10}$~\cite{ExpLaH10-PRL2019, ExpLaH10-Nature2019}, respectively. It was previously pointed out that for LaH$_{10}$, the anharmonic effects on phonons and the quantum ionic zero-point energy are of importance for a proper prediction of $P_c$~\cite{LaH10-anharmonic}. Therefore, the above overestimation of predicted $P_c$ values is likely due to the ignorance of anharmonic and quantum effects~\cite{H3S-anharmonic-1,H3S-anharmonic-2,H3S-nonadiabatic} in the present theory. Nevertheless, we can say that ThH$_{10}$, ThH$_9$, and CeH$_9$ have significantly larger chemical precompressions compared to LaH$_{10}$.

\begin{figure}[h!b]
\centering{ \includegraphics[width=8.0cm]{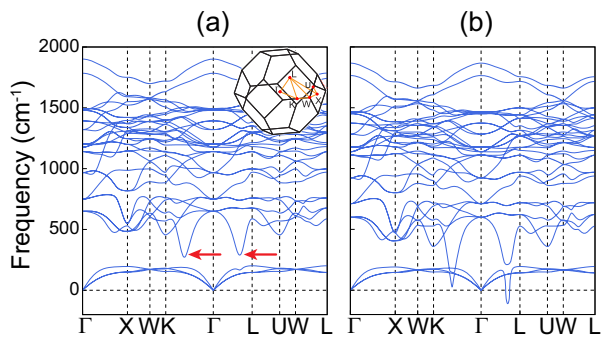} }
\caption{Calculated phonon spectra of ThH$_{10}$ at (a) 130 and (b) 120 GPa. The arrows in (a) indicate the softened phonon modes. The imaginary phonon frequencies appear at 120Gpa along the $\Gamma-L$ line.}
\label{figure:3}
\end{figure}

We next explore the electride-like characteristics of the isolated metal frameworks of ThH$_{10}$, ThH$_9$, CeH$_9$, and LaH$_{10}$. Here, the structure of each metal framework is taken from the optimized structure of the corresponding superhydride. The valence charge densities ${\rho_{\rm M}}$ (without including of semicore electrons) of the metal frameworks of ThH$_{10}$, ThH$_9$, CeH$_9$, and LaH$_{10}$, calculated at an equal pressure of 300 GPa, are displayed in Figs. 4(a), 4(c), 4(e), and 4(g), respectively. It is seen that some electrons detached from metal atoms are localized in the interstitial regions around the $A_1$ and $A_2$ sites. These interstitial excess electrons of the so-called $A_1$ and $A_2$ anions are well confirmed by the corresponding electron localization function (ELF)~\cite{ELF}. Figures. 4(b), 4(d), 4(f), and 4(h) represent the calculated ELFs of the metal frameworks of ThH$_{10}$, ThH$_9$, CeH$_9$, and LaH$_{10}$, respectively. For the Th framework of ThH$_{10}$, the number of electrons $Q_{A_1}$ ($Q_{A_2}$) within the muffin-tin sphere of the $A_1$ ($A_2$) anion is $-$0.281 ($-$0.234)$e$, larger in magnitude than $-$0.204 ($-$0.194)$e$ for the La framework of LaH$_{10}$ [see Figs. 4(a) and 4(g)]. Similarly, as shown in Figs. 4(c) and 4(e), $Q_{A_1}$ ($Q_{A_2}$) in ThH$_9$ is $-$0.124 ($-$0.080)$e$, larger in magnitude than the corresponding value of $-$0.118 ($-$0.072)$e$ in CeH$_9$. Therefore, the metal framework of ThH$_{10}$ (ThH$_9$) exhibits a more electride feature than that of LaH$_{10}$ (CeH$_9$). It is noted that the magnitudes of $Q_{A_1}$ and $Q_{A_2}$ change as a function of pressure (see Figs. S3 and S4 in the Supplemental Material), showing that the electride-like characteristics of metal frameworks are enhanced with increasing pressure. Indeed, the localization of interstitial excess electrons also emerges in compressed alkali metals at high pressures~\cite{hoffman,hosono,Li6P}, in order to reduce Coulomb repulsions arising from the overlap of atomic valence electrons. Such loosely bound anionic electrons residing in the metal frameworks of compressed superhydrides can be easily captured to H atoms, forming H cages with attractive Coulomb interactions between cationic metal and anionic H atoms. It is remarkable that the anionic electrons in H cages are mostly supplied because of the electride nature of metal frameworks at high pressures, rather than due to the different electronegativities of metal and H atoms~\cite{rare-earth-hydride-PRL2017}.

\begin{figure}[htb]
\centering{ \includegraphics[width=8.0cm]{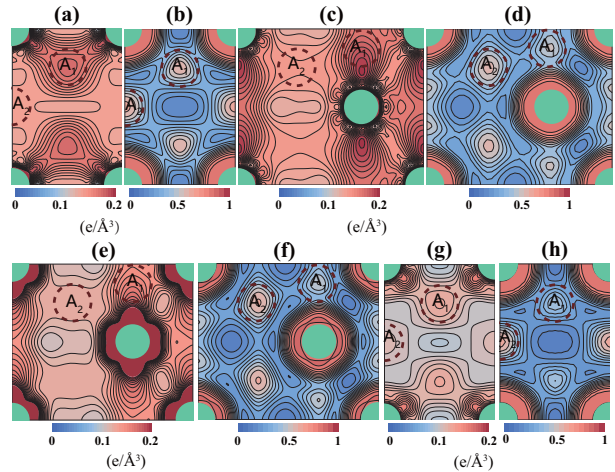} }
\caption{ Calculated valence charge density plot of the metal frameworks of (a) ThH$_{10}$, (c) ThH$_{9}$, (e) CeH$_{9}$, and (g) LaH$_{10}$ at 300 GPa, together with the ELFs of the metal frameworks of (b) ThH$_{10}$, (d) ThH$_{9}$, (f) CeH$_{9}$, and (h) LaH$_{10}$. The charge densities in (a), (c), (e), and (g) are drawn on the (110) plane with a contour spacing of 0.005 $e$/{\AA}$^3$. The ELF in (b), (d), (f), and (h) are drawn with a contour spacing of 0.05. A$_1$ and A$_2$ indicate the two anions in interstitial regions, and the dashed circles represent the muffin-tin spheres around A$_1$ and A$_2$ with the radii of 0.75 (0.60) and 0.75 (0.60) {\AA} in ThH$_{10}$ and LaH$_{10}$ (ThH$_{9}$ and CeH$_{9}$), respectively.}
\label{figure:4}
\end{figure}

Figures 5(a) and 5(b) show the calculated total charge densities of ThH$_{10}$ and ThH$_9$ at 300 GPa, respectively. It is seen that the H atoms in each H cage are bonded to each other with covalent bonds, where each H$-$H bond has a saddle point of charge density at its midpoint, similar to the C$-$C covalent bond in diamond~\cite{diamond}. For ThH$_{10}$ (TH$_9$), the charge densities ${\rho}_{\rm H-H}$ at the midpoints of the H$-$H bonds are 0.911 and 0.729 (0.992, 0.769, and 0.601) $e$/{\AA}$^3$: see the arrows in Figs. 5(a) and 5(b). In order to confirm that the interstitial excess electrons of the Th framework of ThH$_{10}$ (ThH$_9$) are captured to form the H$_{32}$ (H$_{29}$) cages, we calculate the charge densities of the isolated H$_{32}$ (H$_{29}$) cages without Th atoms. Here, the structure of each isolated H cage is taken from the optimized structure of the corresponding superhydride. As shown in Fig. 5(c) [5(d)], we find that ${\rho}_{\rm H-H}$ decreases as 0.742 and 0.621 (0.843, 0.580, and 0.479) $e$/{\AA}$^3$, smaller than those in ThH$_{10}$ (TH$_9$). This indicates that the H$-$H covalent bonds in ThH$_{10}$ and TH$_9$ are strengthened by capturing the interstitial excess electrons of isolated Th frameworks.

\begin{figure}[htb]
\centering{ \includegraphics[width=8.0cm]{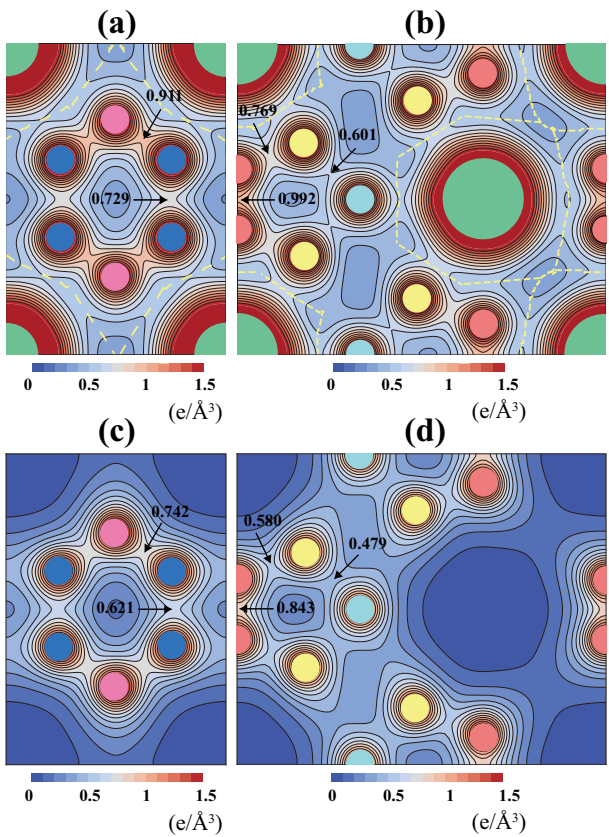} }
\caption{Calculated total charge densities of (a) ThH$_{10}$ and (b) ThH$_{9}$ at 300 GPa, with a contour spacing of 0.1 $e$/{\AA}$^3$. The Bader basins of Th atoms are also drawn in (a) and (b). The calculated charge densities of the isolated H$_{32}$ and H$_{29}$ cages of ThH$_{10}$ and ThH$_{9}$ without Th atoms are displayed in (c) and (d), respectively. The numbers represent the values of ${\rho}_{\rm H-H}$ at the midpoints (marked by the arrows) of the H$-$H bonds. }
\label{figure:4}
\end{figure}

To provide an explanation for the variation of $P_c$ in ThH$_{10}$, ThH$_9$, CeH$_9$, and LaH$_{10}$, we estimate chemical precompression by calculating the attractive Coulomb forces between a metal atom and its surrounding H atoms [see the lower panel in Figs. 1(a) and 1(b)]. This simple estimation is based on the complete screening of the electric field arising from metal atoms within H cages. Using the Bader~\cite{Bader} analysis, we calculate the cationic charge $Q_{\rm M}$ inside the Bader basin [see Figs. 5(a) and 5(b)] of metal atom in each superhydride. For ThH$_{10}$, ThH$_9$, CeH$_9$, and LaH$_{10}$, we obtain $Q_{\rm M}$ values of 1.486, 1.464, 1.199, and 1.036$e$, respectively (see Fig. 6). Assuming that $Q_{\rm M}$ is the point charge at the position of corresponding metal atom and the anionic charge ($-$$Q_{\rm M}$) of H atoms is uniformly distributed on the spherical shell with a radius of $d_{\rm M-H}$, we evaluate the magnitudes of Coulomb forces acting on the H atoms composing the H$_{32}$ or H$_{29}$ cage, and divide it by the surface area of the spherical shell. Figure 6 shows such estimated chemical pressures of ThH$_{10}$, ThH$_9$, CeH$_9$, and LaH$_{10}$ at 300 GPa, with ratios relative to the value of LaH$_{10}$. We find that the chemical pressures of ThH$_{10}$ and ThH$_9$ (CeH$_9$) are about two (one and half) times higher than that of LaH$_{10}$, indicating that the former Group-4 metal hydrides have larger chemical precompressions to attain lower $P_c$ values than the latter Group-3 metal hydride. Considering that the $d_{\rm M-H}$ values of ThH$_{10}$ and ThH$_9$ are close to that of LaH$_{10}$ [see Fig. 2(a)], the higher chemical pressures in Th superhydrides are likely attributed to more cationic and anionic charges compared to LaH$_{10}$. As shown in Fig. 6, the chemical pressures of four superhydrides are well consistent with their variations of $Q_{\rm M}$. It is noted that the estimated chemical pressure of CeH$_9$ is lower than that of isostructural ThH$_9$ at 300 GPa (see Fig. 6), while the predicted value of $P_c$ = 100 GPa for the former is lower than that (110 GPa) for the latter. This inconsistency of chemical precompression and $P_c$ between CeH$_9$ and ThH$_9$ may be due to the delocalized nature of Ce 4$f$ electrons~\cite{hyunsoo}, which could lower $P_c$ via a more hybridization with the H 1$s$ state. Nevertheless, despite their crude simulations, the estimated relative chemical pressures of ThH$_{10}$, ThH$_9$, CeH$_9$, and LaH$_{10}$ are in reasonable agreement with the variation of experimentally measured $P_c$ values~\cite{ExpLaH10-PRL2019, ExpLaH10-Nature2019,ExpThH10-Materials Today2020-Oganov,ExpCeH9-Nat.Commun2019T.Cui,ExpCeH9-Nat.Commun2019-J.F.Lin,ExpCeH9-arXiv2021}.

\begin{figure}[htb]
\centering{ \includegraphics[width=8.0cm]{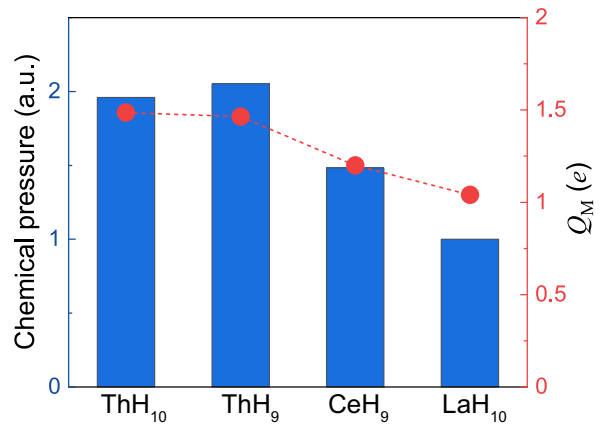} }
\caption{Calculated chemical pressures of ThH$_{10}$, ThH$_{9}$, CeH$_{9}$, and LaH$_{10}$ at 300 GPa. The cationic charges $Q_{\rm M}$ of metal atoms obtained from Bader charge analysis are also given.}
\label{figure:4}
\end{figure}

\section{Summary}

Using first-principles DFT calculations, we have conducted a comparative study of chemical precompressions in experimentally synthesized superhydrides including ThH$_{10}$, ThH$_9$, CeH$_9$, and LaH$_{10}$. We found that these superhydrides form H clathrates by capturing excess electrons in interstitial regions of their isolated fcc- and hcp-metal frameworks. By taking into account the attractive Coulomb interactions between cationic metal atom and its surrounding H atoms, we estimated chemical precompressions in ThH$_{10}$, ThH$_9$, CeH$_9$, and LaH$_{10}$. It was found that ThH$_{10}$, ThH$_9$, and CeH$_9$ have larger chemical precompressions than LaH$_{10}$, consistent with the variation of experimentally measured $P_c$ values~\cite{ExpLaH10-PRL2019, ExpLaH10-Nature2019,ExpThH10-Materials Today2020-Oganov,ExpCeH9-Nat.Commun2019T.Cui,ExpCeH9-Nat.Commun2019-J.F.Lin,ExpCeH9-arXiv2021}. Our findings not only demonstrated that interstitial excess electrons in the metal frameworks of superhydrides play an importnt role in generating the chemical precompression of H cages around metal atoms, but also have important implications for the exploration of new superhydrides which can be synthesized at moderate pressures below ${\sim}$100 GPa.

\section{Supplementary Material}
See Supplemental Material for the lattice constants of ThH$_{10}$, ThH$_{9}$, CeH$_{9}$, and LaH$_{10}$, the phonon spectrum of ThH$_{9}$, and the valence charge densities of the Th frameworks of ThH$_{10}$ and ThH$_{9}$.

\section{Author's contributions}
S. Y., C. W., and S. L. contributed equally to this work.

\section{Acknowledgement}
This work was supported by the National Research Foundation of Korea (NRF) grant funded by the Korean Government (Grants No. 2019R1A2C1002975, No. 2016K1A4A3914691, and No. 2015M3D1A1070609). The calculations were performed by the KISTI Supercomputing Center through the Strategic Support Program (Program No. KSC-2020-CRE-0163) for the supercomputing application research.

\section{DATA AVAILABILITY}
The data that support the findings of this study are available from the corresponding author upon reasonable request.




\vspace{0.6cm}

\noindent $^{*}$ Corresponding author: chojh@hanyang.ac.kr


\end{document}